# Unveiling the nonrelativistic spin current polarization in an altermagnet

Chang Pan,[1,*] Shuai Hu,[2,*] Fanli Yang,[3] Dongchao Yang,[1] Weijia Fan,[1] Zhong Shi,[1] Liqing Pan,[3,†] Shiming Zhou,[1,†] Xuepeng Qiu[1,†]

[1] *School of Physics Science and Engineering, Tongji University, Shanghai 200092, China*

[2] *Institute of Quantum Materials and Devices, School of Electronic and Information Engineering, Tiangong University, Tianjin, 300387, China.*

[3] *Hubei Engineering Research Center of Weak Magnetic-field Detection, College of Science, China Three Gorges University, Yichang, 443002, China*

[*]These authors contribute equally to this work

[†]e-mail: lpan@ctgu.edu.cn, shiming@tongji.edu.cn and xpqiu@tongji.edu.cn

## Abstract

Spin current plays a central role in spintronics for driving exotic spin-dependent phenomena and high-performance device applications. Recently, a magnetic spin Hall effect has been discovered in spin-split antiferromagnets including noncollinear antiferromagnets and altermagnets, allowing the efficient generation of unconventional spin currents even in the absence of spin-orbit coupling. However, although such nonrelativistic spin currents are proposed to have a magnetic origin, the direct connection between the Néel vector and the spin current polarization is still missing. Here, using the altermagnetic $RuO_2$ as a representative example, we unveil the spin current polarization by disentangling the conventional and magnetic spin Hall effects using a technique we developed based on the spin Hall magnetoresistance measurement. The results suggest that the nonrelativistic spin current hosts a polarization very close to the Néel vector. Our work offers unambiguous evidence of the magnetic origin of the nonrelativistic spin current in altermagnetic $RuO_2$, and paves a straightforward route to understand the unconventional spin currents that are crucial in spintronics.

Spintronics exploits the spin degree of freedoms in electronic devices for information processing and storage [1]. The magnetic order parameters of magnetic materials serve as the state variables in spintronic devices, making the manipulation of magnetic states essential for the write-in operation. "Standard" method for this sake relies on a magnetic field, which suffers a significant energy dissipation. Recently, it was found that the electrically generated spin currents can exert spin torques in magnets [2-10], providing an effective way on manipulating the magnetization in spintronic devices.

Conventionally, there are two approaches for generating spin currents and the associated spin torques. The first one utilizes a current perpendicular-to-plane (CPP) device, such as a magnetic tunnel junction (MTJ), where a longitudinal spin current from a ferromagnetic electrode tunnels across the barrier and transfers the spins towards another electrode [11-13]. Such a spin current is polarized by the spin-split Fermi surface of the ferromagnetic electrode, and hence is collinear to the magnetization. The resultant spin-transfer torque (STT) directly competes with the damping of the magnetization, leading to an efficient switching [13]. This approach, however, requires a substantial incubation time to initial the torque dynamics, lowering the switching speed [14,15]. Another approach employs the phenomena associated with spin-orbit coupling (SOC), such as spin Hall effect (SHE) [10,16] and Rashba-Edelstein effect (REE) [9,17,18], generating a transverse spin current from a spin source layer or an interface and exerting the torques on the adjacent magnetic layer. Although such spin-orbit torque (SOT) in the associated current in-plane (CIP) devices shows the advantages such as the absence of read–write–breakdown voltage interferences, the difficulty to realize a deterministic switching of perpendicular magnetization limit its application [9,10,19].

Recent findings have shown that some antiferromagnets, including the noncollinear antiferromagnets and altermagnets, host momentum-dependent spin splitting even in the absence of SOC [20-31]. The anisotropic spin distribution on Fermi surface of these spin-split antiferromagnets results in unconventional transverse spin currents [20,25,32-42]. This nonrelativistic spin current is time-reversal-odd and can be reversed by switching the Néel vector [20,32,40]. Such a phenomenon has been dubbed magnetic spin Hall effect (MSHE) [20,33] or time-reversal-odd ($\hat{T}$-odd) [41], to distinguish it from the conventional time-reversal-even ($\hat{T}$-even) SHE due to SOC. Since the $\hat{T}$-odd nonrelativistic spin current is polarized collinear to the Néel vector, it hosts unconventional spin polarization incapable of the $\hat{T}$-even spin current, and hence can be used to realize efficient and deterministic switching in CIP spin torque devices [34,36,41].

So far, the unconventional spin polarization of the $\hat{T}$-odd spin currents have been observed in many split antiferromagnets, including Mn$_3$AN (A=Ga, Sn) [34,43], Mn$_3$X (X=Sn, Pt) [36,37,44], and RuO$_2$ [40-42,45-47]. However, the existing experiments are not sufficient to connect the spin current polarizations to the Néel vector of these antiferromagnets. One possible reason is that the $\hat{T}$-odd and $\hat{T}$-even spin currents are intertwined [37] and hence a real polarization of the $\hat{T}$-odd spin current is difficult to derive. A typical example is rutile RuO$_2$, a canonical candidate of recently proposed altermagnet, which is widely employed to demonstrate spin-dependent transport properties such as anomalous Hall effect [37,48,49], unconventional spin current [32,40-42,50], tunneling magnetoresistance [51-54] and spin-transfer torque [52] in antiferromagnetic tunnel junctions [55]. In contrast to the theoretical prediction [32], the experimentally measured spin current polarization of RuO$_2$ shows strong deviation from the Néel vector direction [41]. Recently, several theoretical and experimental [56-58] works suggest a nonmagnetic ground state of RuO$_2$, bringing further question about the origins of the above-mentioned spin-dependent transport properties.

In this work, we try to solve the puzzle of the nonrelativistic spin current polarization by disentangling the $\hat{T}$-odd and $\hat{T}$-even spin currents. Considering RuO$_2$ as a representative example, we identify spin current polarization in the RuO$_2$/CoFeB heterojunction at different temperature using a technique we developed based on the spin Hall magnetoresistance (SMR) measurement [59,60], and decompose its $\hat{T}$-odd and $\hat{T}$-even components based on their disparate temperature dependence. The results suggest the $\hat{T}$-odd spin current is dominant and host the polarization very close to Néel vector direction at low temperature, indicating its magnetic origin. We also find the conventional $\hat{T}$-even spin current is very strong and is dominant at high temperature.

The $\hat{T}$-odd and $\hat{T}$-even spin currents arise from different origins. In a collinear magnetic metal under the nonrelativistic limit, the conductivity $\sigma_{ba}$ can be decomposed into that contributions from the spin-up (↑) and spin-down (↓) Fermi surfaces, expressed as $\sigma_{ba} = \sigma_{ba}^{\uparrow} + \sigma_{ba}^{\downarrow}$, where $a$ and $b$ are applied electric field and the generated charge current directions, respectively. The $\hat{T}$-odd spin conductivity is then $\sigma_{ba}^{s,odd} = \sigma_{ba}^{\uparrow} - \sigma_{ba}^{\downarrow}$ (Fig. 1(a)), where the spin polarization $s$ is parallel to the magnetic order parameter [32]. On the other hand, the $\hat{T}$-even spin conductivity $\sigma_{ba}^{s,even}$ is associated with SOC and contributed by the spin-Berry curvature within the band gaps [16], where $s$ is determined by the crystal symmetry operations (Fig. 1(b)). Since exchange splitting is usually much stronger than SOC, the $\hat{T}$-odd spin current is expected to be much stronger than the $\hat{T}$-even one.

RuO$_2$ is a rutile antiferromagnetic candidate widely used to demonstrate the spin-dependent transport properties associated with the altermagnetic spin splitting [32,40-42,50] and the C-type antiferromagnetic stacking [52] useful for spintronic devices such as antiferromagnetic tunnel junctions [51-55]. Especially, it was predicted to host notable $\hat{T}$-odd spin current [32], and both the conventional component $\sigma_{zx}^{y,odd}$ and the unconventional $\sigma_{zx}^{x,odd}$ and $\sigma_{zx}^{z,odd}$ of spin conductivity have been found in low symmetric orientated (101) and (110) RuO$_2$ films [40-42], which are considered to be associated with the $\hat{T}$-odd spin current. However, the spin current polarization derived based on these components is not consistent with the Néel vector direction. This is possibly due to the $\hat{T}$-even spin conductivity, which also hosts these components [41] and might be not negligible in contrast to the expectation. As a result, the intertwined $\hat{T}$-odd and $\hat{T}$-even spin currents lead to an anomalous spin polarization [37].

Since $\sigma_{ba}^{s,odd}$ and $\sigma_{ba}$ are strongly related, they should have the same temperature dependence, i.e. $\propto T^{-n}$ (Fig. 1(c)). On the other hand, the $\sigma_{ba}^{s,even}$ is an intrinsic properties of band structure and independent of scattering, and hence is expected to be weakly dependent on temperature [41]. Therefore, it is possible to disentangle $\sigma_{ba}^{s,odd}$ and $\sigma_{ba}^{s,even}$ based on their different temperature dependence and unveil the ultimate nonrelativistic spin polarization of RuO$_2$.

Here we develop a technique based on SMR measurement to directly measure the spin current polarization of RuO$_2$. SMR is a transport phenomenon associated with charge-spin conversion and its inverse effect. An in-plane charge current is first converted to an out-of-plane spin current with a polarization $\boldsymbol{s}$, which is reflected or absorbed by the adjacent ferromagnetic layer depending on the ferromagnetic magnetization $\boldsymbol{m}$. The reflected spin current, with the intensity proportional to $(\boldsymbol{m} \cdot \boldsymbol{s})^2$, is then converted back to an in-plane charge current by the inverse effect in addition to the original charge current, resulting in a variation of the overall resistance. Therefore, one can in principle derive the $\boldsymbol{s}$ of a spin current from the SMR measurement. This technique is especially suitable for investigating the temperature dependence of $\boldsymbol{s}$, since SMR measurement can be performed with a small DC current and hence avoids the substantial Joule heating which influences the device temperature and enhances the scattering.

We sequentially grew (101)-oriented RuO$_2$ and amorphous CoFeB on Al$_2$O$_3$ (1$\bar{1}$02) substrates (details in Supplementary Note 1). Here, current is along the [010] crystallographic direction, and the current direction was defined as $\boldsymbol{x}$, and the normal direction as $\boldsymbol{z}$. During the measurement, a 4 T external magnetic field was rotated in the $\boldsymbol{y}$-$\boldsymbol{z}$ plane, as illustrated in Fig. 2(a), while a 100 μA DC current along the $\boldsymbol{x}$ direction and measured the longitudinal magnetoresistance ($R$) using the four-point method. The fitting blue line in Fig. 2(d) shows the SMR of the RuO$_2$/CoFeB sample at 5 K. Unlike traditional SMR that majorly follows a $\Delta R\cos(2\beta)$ angular dependence [60], the SMR in Fig. 2(d) (blue line) takes the form $\Delta R\cos(2(\beta-\Delta\beta))$, where $\beta$ is the angle between the magnetic field and $\boldsymbol{z}$, $\Delta\beta$ is the phase shift, and $\Delta R$ is the SMR magnitude defined as a half of the difference between the maximum and the minimum value of resistance $R(\beta)$. To eliminate instrumental errors, we deposited Pt/CoFeB on the same Al$_2$O$_3$ substrate (Supplementary Note 2) as Fig. 2(c) and measured its SMR to calibrate the angle when measuring SMR of RuO$_2$/CoFeB. In Pt/CoFeB, the spin polarization direction is along *y*, so when *m*//*y*, the spin current reflected by the FM layer is maximized, leading to the minimum magnetoresistance due to the current induced by the inverse spin Hall effect (ISHE) in Pt, as shown in Fig. 2(d) dark line. When we apply a current along the [010] direction in RuO$_2$(101), the spin current polarization direction has both *y* and *z* components, the vector sum of their polarization is denoted as *s*. Therefore, the SMR reaches its minimum when *m*//*s*, with *m* not parallel to *y*. We also measured SMR with the current along [$\bar{1}$01], where the magnetoresistance is minimal when *m*//*y*, plotted in Fig. 2(d) (yellow line), indicating no *z* component in the spin polarization, consistent with theoretical predictions.

Figure 3(a,b) show the SMR of RuO$_2$/CoFeB at various temperatures with the current along [$\bar{1}$01] and [010] directions. Negligible magnetoresistances of RuO$_2$ and CoFeB rule out effects of themselves, shown in Supplementary Note 6. We find Δ$\beta$ remains zero when current along [$\bar{1}$01] (Fig. 3(a,c)). This is because the charge current is parallel to a mirror plane M$_{[010]}$ in this case, which only allows the generation of a *y*-polarized spin current. Therefore, it is difficult to reveal the nonrelativistic contribution of the SMR based on the measurements when current along [$\bar{1}$01]. On the other hand, we find sizable Δ$\beta$ when the current is along [010] (Fig. 3(b,c)), clearly indicating the existence of unconventional spin current polarization. Both Δ$\beta$ and Δ$R$ decreases at high temperature (Fig. 3(c,d)), implying the $\hat{T}$-odd SHC decreases and $\hat{T}$-even SHC becomes dominant at high temperature.

Next, we use the SMR results to decompose the $\hat{T}$-odd and $\hat{T}$-even SHC based on their temperature dependence. Assuming zero longitudinal spin absorption and a transparent RuO$_2$/CoFeB interface, the SMR results of RuO$_2$ fit the following equation [61]:

$$\frac{\Delta R}{R} = -\theta_{SH}^2 \tanh(\varepsilon)\frac{\lambda_{sf}}{t_{AF}}[1 - \frac{1}{\cosh(2\varepsilon)}], \tag{1}$$

where $\theta_{SH}$ is the spin Hall angle (Fig. S1 of Supplementary Note 3), $\lambda_{sf}$ is the spin diffusion length of 12.2 nm [62], $t_{AF}$ is the thickness of RuO$_2$, and $\varepsilon = t_{AF}/2\lambda_{sf}$.

The spin Hall conductivity $\sigma_{zx} = \theta_{SH}\sigma_{xx}\hbar/2e$, where $\sigma_{xx} = 1/\rho_{xx}$. ($\rho_{xx}$-$T$ curve in Fig. S2 of Supplementary Note 3) Since we know the angle $\Delta\beta$, we can decompose the SHC of RuO$_2$ into $\sigma_{zx}^{z} = \sigma_{zx} \cdot \sin\Delta\beta$ and $\sigma_{zx}^{z} = \sigma_{zx} \cdot \cos\Delta\beta$ in the orthogonal $z$ and $y$ directions, as shown in Fig. 4(a). The $\hat{T}$-odd SHC is proportional to the electron lifetime $\tau$, which follows $\tau = \tau_0(1/T)$ between 150 K and 300 K, making $\hat{T}$-odd SHC proportional to $1/T$ within this temperature range. Given that the temperature dependence of $\hat{T}$-even SHC is negligible, we can fit within this temperature range using the relation $\sigma_{zx}^{z} = A(1/T) + B$, as shown by the blue line in Fig. 4(b), where $\sigma_{zx}^{z,\text{even}} = B$ is denoted by the purple line in Fig. 4(b). We then obtain $\sigma_{zx}^{z,\text{odd}} = \sigma_{zx}^{z} - \sigma_{zx}^{z,\text{even}}$ over the entire temperature range, shown by the red line in Fig. 4(b). For $\sigma_{zx}^{y}$, we fit its temperature dependence similarly, as shown in Fig. 4(c).

We take $\arctan(\sigma_{zx}^{y,\text{odd}}/\sigma_{zx}^{z,\text{odd}})$ as $\theta_s^{\text{odd}}$, the angle between the $\hat{T}$-odd spin current polarization direction and $z$. $\theta_s^{\text{odd}}$ at different temperatures is plotted in Fig. 4(d). In (101)-oriented RuO$_2$, considering lattice parameter a = b = 4.59 Å and c = 3.12 Å, the angle between the Néel vector and $z$ is $\theta_N = \arctan(c/a) = 35°$. In Fig. 4(d) it can be seen that within the range of 5 K to 360 K, the $\hat{T}$-odd spin current polarization direction is roughly parallel to the Néel vector.

Our results imply two interesting facts: First, the $\hat{T}$-odd spin current is indeed have a magnetic origin and polarized along the Néel vector in RuO$_2$, despite the nonmagnetic ground state suggested by many recent theoretical and experimental works [56-58]. This may indicate the unavoidable factors induced by film growth, such as strain, oxygen vacancies, interfacial charge transfer, anti-site defects, etc., play important roles in stabilizing the magnetism. Second, although it is generally believed the SOC is a weak effect compared to the magnetic exchange splitting, the $\hat{T}$-even $y$-polarized spin current driven by SOC is robust and even much stronger than $\hat{T}$-odd one driven by magnetism, due to the latter being largely suppressed by the scattering at high temperature. At last, we note the method we used is valid not only for altermagnets such as RuO$_2$, but also for other unconventional antiferromagnets such as noncollinear antiferromagnets [34,36,37,43,44]. Moreover, it can be efficient to unveil the unconventional polarization of spin currents with other exotic mechanisms, such as the composition gradient [63-66], symmetry control [67,68], interface engineering [69], etc.

In conclusion, using the technique we developed based on SMR measurements, we have disentangled the $\hat{T}$-odd and $\hat{T}$-even spin current of altermagnetic RuO$_2$, and found the nonrelativistic $\hat{T}$-odd spin current is polarized along the Néel vector. Our results unambiguously reveal the magnetic origin of the nonrelativistic spin current in altermagnetic RuO$_2$, and pave a straightforward route to understand the unconventional spin currents that are crucial in spintronics.

**Acknowledgments**

This work was supported by the National Key R&D Program of China (Grand No. 2022YFA1204002), the National Natural Science Foundation of China (Grant Nos. 52371246, 12274258, 12274323, 52271188, 12074285), the International Science and Technology Cooperation Program under the 2023 Shanghai Action Plan for Science, Technology and Innovation (Grant No. 23520711200), the Natural Science Foundation of Shanghai (Grant No. 23ZR1466800), Open Fund of the State Key Laboratory of Spintronics Devices and Technologies (Grant No. SPL-2412).

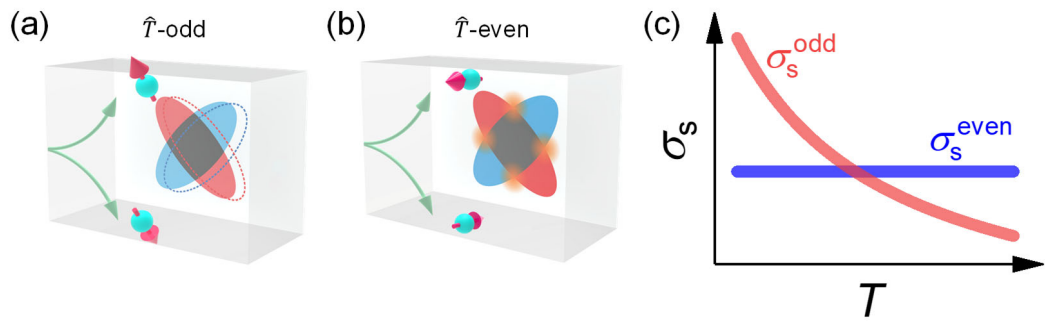

**FIG. 1: (a,b)** Schematics of $\hat{T}$-odd and $\hat{T}$-even charge-spin conversion in altermagnetic $RuO_2$ due to nonrelativistic spin splitting (**a**) and conventional spin Hall effect (**b**). **(c)** A schematic of temperature dependence of $\hat{T}$-odd and $\hat{T}$-even SHC with different temperature dependence.

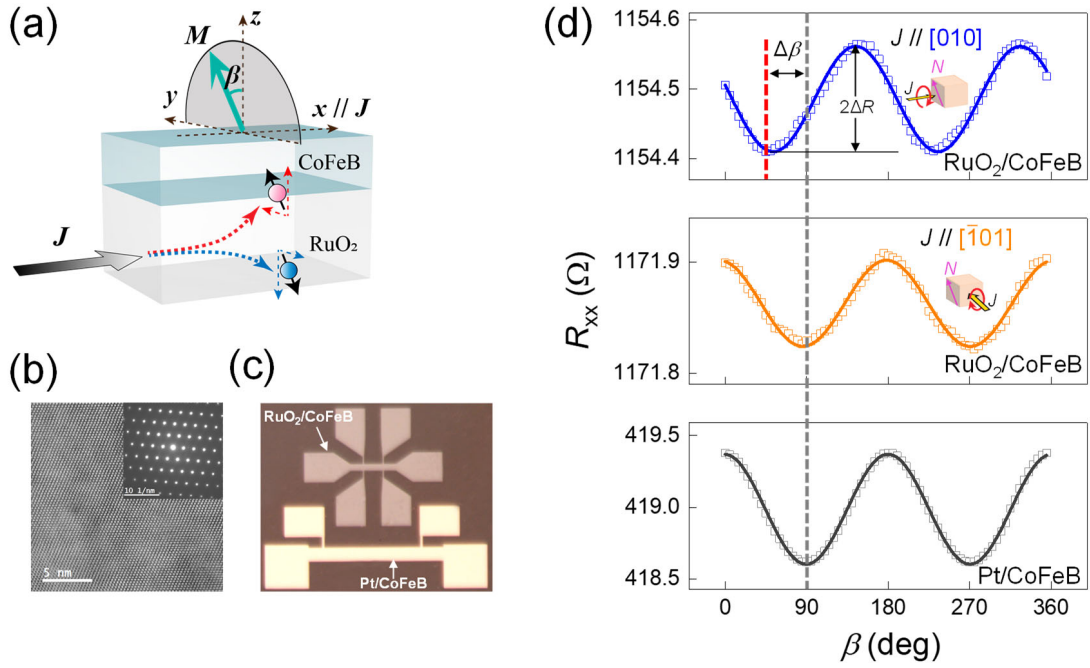

**FIG. 2:** The spin polarization direction of RuO$_2$ is extracted by SMR. **(a)** SMR measurement scheme of RuO$_2$/CoFeB heterojunction: magnetoresistance measured while scanning a 4 T magnetic field in a plane perpendicular to the electric field, and $\beta$ is the angle between the magnetic field and *z*. **(b)** Cross-sectional scanning transmission electron microscopy (STEM) image of RuO$_2$(101). Inset: fast Fourier transform (FFT) of the STEM image. **(c)** Optical image of the RuO$_2$/CoFeB (upper) and Pt/CoFeB (lower) device using for SMR measurements, in the former current direction along RuO$_2$[010] and current in the latter is parallel to the former. **(d)** SMR in RuO$_2$/CoFeB with currents along [010] (blue point) and [$\bar{1}$01] (yellow point) and SMR of Pt/CoFeB (dark point) as a reference sample.

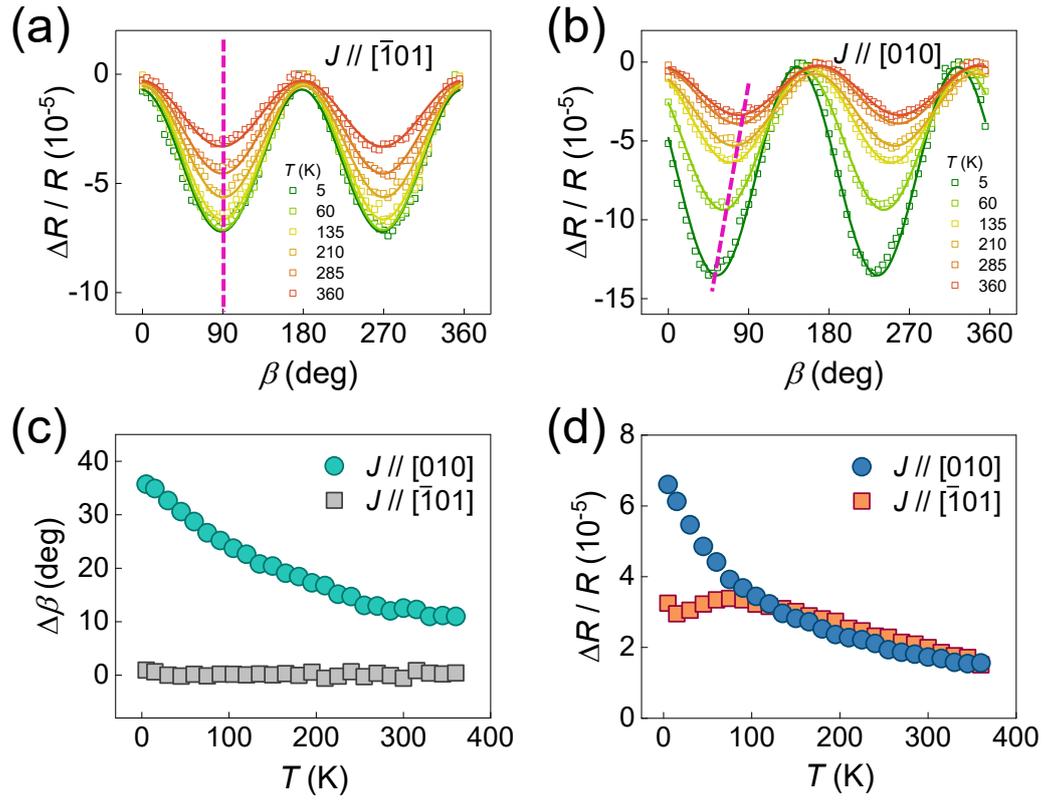

**FIG. 3:** Temperature dependence of SMR. **(a,b)** The SMR at different temperatures for the current along the $[\bar{1}01]$ (a) and $[010]$ (b) directions. **(c,d)** The temperature dependence of the $\Delta\beta$ (c) and $\Delta R$ (d) derived from the SMR shown in (a) and (b).

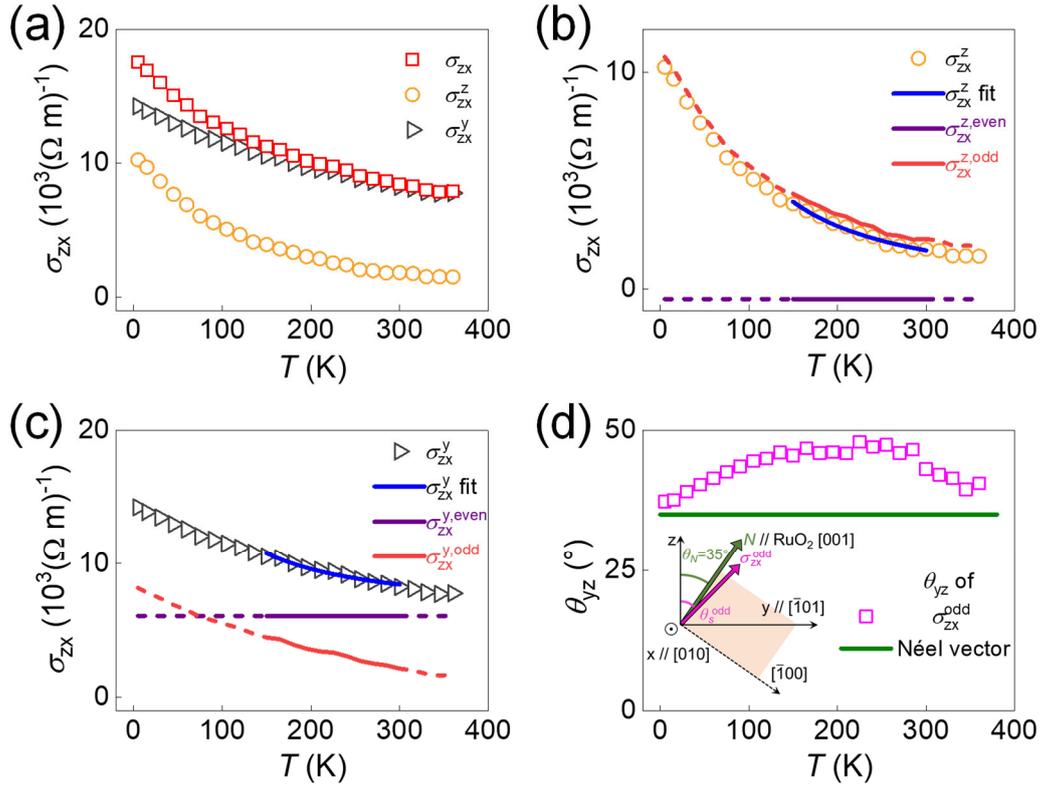

**FIG. 4:** Decomposition of the total SHC into $\hat{T}$-odd and $\hat{T}$-even components via temperature dependence. **(a)** The total, *y*- and *z*-polarized SHC of RuO$_2$ derived from SMR measurement. **(b,c)** The $\hat{T}$-odd and $\hat{T}$-even components of *z*-polarized (b) and *y*-polarized (c) SHC decomposed by fitting the temperature dependence of the SHC. **(d)** The $\hat{T}$-odd spin current polarization indicated by the angle $\theta_s^{odd} = \arctan(\sigma_{zx}^{y,odd}/\sigma_{zx}^{z,odd})$ as a function of temperature.